\documentclass{aastex}
\shorttitle{Toroidal L and H equilibria with axisymmetric rotations}
\shortauthors{Tsui}
\begin{document}
\title{Toroidal L and H equilibria with axisymmetric rotations}
\author{K.H. Tsui}
\affil{Instituto de F\'{i}sica - Universidade Federal Fluminense,
\\Campus da Praia Vermelha, Av. General Milton Tavares de Souza s/n,
\\Gragoat\'{a}, 24.210-346, Niter\'{o}i, Rio de Janeiro, Brasil.}
\email{tsui$@$if.uff.br}
\date{}
\pagestyle{myheadings}
\baselineskip 24pt

\begin{abstract}

Axisymmetric toroidal equilibria with toroidal and poloidal rotations
 are solved with a specific set of source functions.
 The two independent solutions are associated to L and H modes.
 The L/H transition is regarded as a bifurcation
 from one equilibrium configuration to another,
 under strong external heating and pellet injection
 to shape temperature and density profiles.
 Because of the steep edge gradient of the H solution,
 large static radial electric field, zonal flow,
 and improved confinements, come as consequences,
 not causes, of the H mode.

\end{abstract}

\vspace{2.0cm}
\renewcommand{\thesection}{\Roman{section}}

\maketitle
\newpage
\section{INTRODUCTION}

The discovery of the H mode with enhanced confinement
 in ASDEX [1] had openned a new age of tokamak fusion research.
 Extensive experimental works were performed
 to identify the key procedures and signatures
 that carried the L to the H confinement mode [2,3].
 These procedures included pellet injection for density enhancement,
 neutral beam or radio frequency heating for temperature
 and electric conductivity profiling, etc [4,5].
 The experimental signatures of L/H transition
 were a drastic reduction of the $D_{\alpha}$ hydrogen emission
 and a sudden decrease of the plasma floating potential.
 These were accompanied by a good enhancement
 of plasma density and energy confinements,
 plus a steepening of the edge plasma profile
 and a reduction of magnetohydrodynamic (MHD) activities [6,7].
 Observationally, the L/H transition was accompanied
 by an increase in the toroidal and poloidal rotations [3].
 Nevertheless, it was not clear that this velocity increase
 was the cause or was the consequence of the transition.

The L/H transition is believed to be caused
 by divertor shaping of the edge plasma,
 under high levels of heating power.
 Action of the divertor plus the edge gradient
 pump up the static radial electric field
 that drives a zonal flow near the
 edge rational magnetic surface [8,9].
 This zonal flow is thought bo cause a transport barrier
 that enhances density and energy confinements.
 The plasma is thought to self-organize gradually
 under this scenario to reach the H mode [10].
 Here, we take a different approach
 to view the L/H transition.
 We consider the toroidal and poloidal rotations
 on the equilibrium scaling, not transport scaling,
 as equilibrium parameters.
 We solve for rotational equilibria,
 under specific source functions,
 in spherical coordinates by seeking toroidal solutions [11].
 There are two independent solutions
 for rotational equilibrium,
 and we associate them to the L and H modes.
 Under this equilibrium configuration approach,
 the transition is seen as a bifurcation
 from one equilibrium to another,
 under external drivings
 such as pellet injection for plasma density
 and strong heating for temperature profiling.
 Because of the steep edge gradient of the H mode,
 the large static radial electric field,
 the zonal flow, and the associated confinements,
 come as the natural consequences, not causes,
 of the H mode solution.

\newpage
\section{ROTATIONAL EQUILIBRIA}

To discuss rotational equilibria,
 we begin with the time-dependent MHD equations

\begin{eqnarray}
\label{eqno1}
{\partial\rho\over\partial t}+\nabla\cdot(\rho\vec v)\,
 =\,0\,\,\,,
\\
\label{eqno2}
\rho\{{\partial\vec v\over\partial t}
 +(\vec v\cdot\nabla)\vec v\}\,
 =\,\vec J\times\vec B-\nabla p\,\,\,,
\\
\label{eqno3}
{\partial\vec B\over\partial t}\,
 =\,-\nabla\times\vec E\,
 =\,\nabla\times(\vec v\times\vec B)\,\,\,,
\\
\label{eqno4}
\nabla\times\vec B\,=\,\mu\vec J\,\,\,,
\\
\label{eqno5}
\nabla\cdot\vec B\,=\,0\,\,\,,
\\
\label{eqno6}
p\,=\,\rho v_{s}^{2}\,\,\,.
\end{eqnarray}

\noindent Here, $\rho$ is the mass density,
 $p$ is the plasma pressure,
 $\vec v$ is the bulk velocity,
 $\vec J$ is the current density,
 $\vec B$ is the magnetic field,
 $v_{s}$ is the ion acoustic speed,
 $\mu$ is the free space permeability.
 With axisymmetry, the magnetic field and the current density
 can be represented by two scalar functions
 in standard spherical coordinates

\begin{eqnarray}
\label{eqno7}
\vec B\,
 =\,A_{0}(\nabla P\times\nabla\phi+Q\nabla\phi)\,
 =\,{A_{0}\over r\sin\theta}
 \{+{1\over r}{\partial P\over\partial\theta},
   -{\partial P\over\partial r},
   +Q\}\,\,\,,
\\
\label{eqno8}
\mu\vec J\,
 =\,{A_{0}\over r\sin\theta}
 \{+{1\over r}{\partial Q\over\partial\theta},
   -{\partial Q\over\partial r},
   -{\partial^2 P\over\partial r^2}
   -{1\over r^2}\sin\theta
   {\partial\over\partial\theta}
   ({1\over\sin\theta}{\partial P\over\partial\theta})\}
   \,\,\,.
\end{eqnarray}

\noindent Here, $A_{0}$ carries the physical dimension
 of poloidal magnetic flux
 such that $P$ is a dimensionless function.
 Also, we can write the axisymmetric
 poloidal and toroidal rotations as

\begin{eqnarray}
\label{eqno9}
\vec v\,
 =\,A'_{0}(\nabla P'\times\nabla\phi+Q'\nabla\phi)\,\,\,.
\end{eqnarray}

\noindent Likewise, $A'_{0}$ carries the physical dimension
 of poloidal velocity flux
 such that $P'$ is a dimensionless function.
 With axisymmetry and incompressible fluid condition,
 $\nabla\cdot\vec v=0$, steady state in Eq.~\ref{eqno1} requires
 
\begin{eqnarray}
\nonumber
\nabla\rho\cdot\vec v\,
 =\,A'_{0}\nabla\rho\cdot(\nabla P'\times\nabla\phi)\,
 =\,A'_{0}(\nabla\rho\times\nabla P')\cdot\nabla\phi\,
 =\,0\,\,\,,
\\
\label{eqno10}
\rho\,=\,\rho_{0}\rho(P')\,\,\,,
\end{eqnarray}

\noindent which requires the poloidal velocity
 and the mass density have the same level contours.
 Here, $\rho_{0}$ carries the dimension and amplitude of mass density,
 and $\rho(P')$ is a dimensionless function.
 As for Eq.~\ref{eqno3}, by Eq.~\ref{eqno7} and Eq.~\ref{eqno9},
 we note that $\vec v\times\vec B$ would be null
 and steady state in Eq.~\ref{eqno3} would be warrented with

\begin{mathletters}
\begin{eqnarray}
\label{eqno11a}
P'\,=\,\alpha P\,\,\,,
\\
\label{eqno11b}
Q'\,=\,\alpha Q\,\,\,.
\end{eqnarray}
\end{mathletters}

\noindent As a result, the velocity field and the magnetic field
 are parallel, generating an emf-free velocity field

\begin{eqnarray}
\label{eqno12}
\vec v\,=\,\alpha {A'_{0}\over A_{0}}\vec B\,
 =\,g\vec B\,\,\,.
\end{eqnarray}

\noindent With Eq.~\ref{eqno4} and Eq.~\ref{eqno6},
 toroidal plasma equilibria of Eq.~\ref{eqno2}
 with full axisymmetric rotations are described by

\begin{eqnarray}
\label{eqno13}
(1-\mu\rho_{0}g^{2}\rho(P))(\nabla\times\vec B)\times\vec B
 -\mu\rho_{0}v_{s}^{2}\nabla\rho(P)\,
 =\,{1\over 2}\mu\rho_{0}g^{2}\rho(P)\nabla B^{2}\,
 =\,0\,\,\,.
\end{eqnarray}

\noindent We consider the rotational scalar pressure
 be much smaller than the plasma pressure,
 thereby giving the second equality in the above equation.
 
\newpage
\section{ROTATIONAL GRAD-SHAFRANOV EQUATION}
 
We seek to solve Eq.~\ref{eqno13} for toroidal solutions.
 This equation renders three components.
 By axisymmetry, the $\phi$ component
 contains only the magnetic force, and it is

\begin{mathletters}
\begin{eqnarray}
\label{eqno14a}
{\partial P\over\partial r}
 {\partial Q\over\partial\theta}
 -{\partial P\over\partial\theta}
 {\partial Q\over\partial r}\,
 =\,0\,\,\,,
\\
\label{eqno14b}
Q(r,\theta)\,=\,Q(P(r,\theta))\,\,\,.
\end{eqnarray}
\end{mathletters}

\noindent As for the $\theta$ component, it reads

\begin{equation}
\label{eqno15}
A_{0}^{2}\{{\partial^2 P\over\partial r^2}
 +{1\over r^2}\sin\theta{\partial\over\partial\theta}
 ({1\over\sin\theta}{\partial P\over\partial\theta})
 +{1\over 2}{\partial Q^{2}\over\partial P}\}\,
 =\,+({v_{s}\over g})^{2}r^2\sin^2\theta
 {\partial\over\partial P}\ln (1-\mu\rho_{0}g^{2}\rho(P))
 \,\,\,.
\end{equation}

\noindent This equation is the rotational counterpart
 of the Grad-Shafranov equation
 of axisymmetric toroidal plasma equilibrium,
 represented in spherical coordinate system.
 The three terms on the left side
 represent the nonlinear force-free field
 with $\mu\vec J=K(P)\vec B$,
 where $K(P)={\partial Q/\partial P}$ is a scalar function.
 This can be verified from Eq.~\ref{eqno7} and Eq.~\ref{eqno8}
 when we impose $\mu\vec J=K(P)\vec B$.
 In particular, we would have the linear force-free field
 should we take $Q^{2}(P)=(aP)^{2}$ with constant $K(P)=a$.
 The term on the right side is the plasma pressure balance.
 The magnetic function $Q^{2}(P)$ and the mass density $\rho(P)$
 are source functions that need to be specified.
 This second order partial differential equation
 has two independent solutions.
 Finally, the $r$ component of Eq.~\ref{eqno13} reads

\begin{eqnarray}
\nonumber
(1-\mu\rho_{0}g^{2}\rho(P)){\partial P\over\partial r}
 A_{0}^{2}\{{\partial^2 P\over\partial r^2}
 +{1\over r^2}\sin\theta{\partial\over\partial\theta}
 ({1\over\sin\theta}{\partial P\over\partial\theta})
 +{1\over 2}{\partial Q^{2}\over\partial P}\}\,
\\
\label{eqno16}
 =\,+({v_{s}\over g})^{2}r^2\sin^2\theta
 {\partial\over\partial r}(1-\mu\rho_{0}g^{2}\rho(P))
 \,\,\,.
\end{eqnarray}

\noindent Comparing Eq.~\ref{eqno16} to Eq.~\ref{eqno15},
 we note that these two equations are identical.
 The $r$ component is simply the self-consistent
 condition of the $\theta$ component.

To solve Eq.~\ref{eqno15} analytically, we take the source
 functions as
 
\begin{mathletters}
\begin{eqnarray}
\label{eqno17a}
Q^2(P)\,=\,a^2P^2+Q^{2}_{0}\,\,\,,
\\
\label{eqno17b}
\ln (1-\mu\rho_{0}g^{2}\rho(P))\,=\,-P{^q}\,\,\,,
\\
\nonumber
\mu\rho_{0}g^{2}\rho(P)\,=\,1-e^{-P{^q}}\,\,\,.
\end{eqnarray}
\end{mathletters}
 
\noindent Writing $P(r,\theta)=R(r)\Theta(\theta)$,
 the rotational Grad-Shafranov equation reads

\begin{eqnarray}
\nonumber
r^{2}{1\over R}{\partial^2 R\over\partial r^2}
 +(ar)^{2}
 +{1\over\Theta}\sin\theta{\partial\over\partial\theta}
 ({1\over\sin\theta}{\partial\Theta\over\partial\theta})\,
\\
\label{eqno18}
 =\,-{v_{s}^{2}\over A_{0}^{2}g^{2}}q(R\Theta)^{q-2}
 r^{4}\sin^2\theta\,
 =\,-{1\over\alpha^{2}}{v_{s}^{2}\over A_{0}^{'2}}q(R\Theta)^{q-2}
 r^{4}\sin^2\theta\,\,\,.
\end{eqnarray}

\noindent The variables of this equation could be separated
 by taking $q=1$ to give

\begin{mathletters} 
\begin{eqnarray}
\label{eqno19a}
(1-x^2){d^2\Theta(x)\over dx^2}
 +n(n+1)\Theta(x)\,=\,0\,\,\,,
\\
\label{eqno19b}
r^{2}{d^2R\over dr^2}
 +[(ar)^{2}-n(n+1)]R\,
 =\,-{1\over\alpha^{2}}{v_{s}^{2}\over (A'_{0}a^{2})^{2}}(ar)^{4}
 {(1-x^2)\over\Theta}\,
 =\,-A_{1}(ar)^{4}{(1-x^2)\over\Theta}\,\,\,,
\end{eqnarray}
\end{mathletters}

\noindent where we have denoted $x=\cos\theta$,
 and used $n(n+1)$ as the separation constant.
 The factor $v_{s}^{2}/(A'_{0}a^{2})^{2}$
 is proportional to $v_{s}^{2}/v_{pol}^{2}$,
 where $v_{pol}$ is the average poloidal rotational velocity.
 The first equation gives

\begin{equation}
\label{eqno20}
\Theta(x)\,=\,(1-x^2){dP_{n}(x)\over dx}\,
 =\,(1-x^2)\,\,\,,
\end{equation}

\noindent where $P_{n}(x)$ is the Legendre polynomial.
 We have taken $n=1$ to get the second equality.
 As for the second equation, the $\theta$ dependent part
 on the right side disappears by having $n=1$.
 The solution is given by $R(r)=R_{0}(r)+R_{1}(r)$,
 where $R_{0}(r)$ and $R_{1}(r)$
 are the homogeneous and particular solutions.
 The homogeneous solution is described by

\begin{equation}
\label{eqno21}
R_{0}(r)\,=\,arj_{n}(ar)+\lambda_{0} ary_{n}(ar)\,\,\,,
\end{equation}

\noindent where $j_{n}(z)$ and $y_{n}(z)$
 are the oscillating spherical Bessel functions,
 and $\lambda_{0}$ is a constant.
 Together with $A_{0}$ defined in Eq.~\ref{eqno7},
 there are two constants for $R_{0}(r)$.
 As for the particular solution, we have

\begin{eqnarray}
\nonumber
z^{2}{d^2R_{1}\over dz^2}+[z^{2}-n(n+1)]R_{1}\,
 =\,-A_{1}z^{4}\,\,\,,
\\
\label{eqno22}
R_{1}(r)\,=\,-A_{1}(ar)^{2}\,=\,-A_{1}z^{2}\,\,\,.
\end{eqnarray}

\noindent We note that the homogeneous solutions,
 $R_{0}(r)$ and $\Theta(x)$,
 correspond to the linear or nonlinear force-free solutions
 of the left side of Eq.~\ref{eqno15}.
 The plasma pressure term on the right side
 appears only in the particular solution, $R_{1}(r)$,
 that keeps the pressure balance.
 The homogeneous radial solution
 is an oscillating function in $z=ar$,
 which has sucessive maxima,
 and the homogeneous meridian solution
 has a lobe peaked at $x=0$.
 The superposition of the particular radial solution
 only slightly modifies the homogeneous solutions.
 We could use the region between $z=0$
 and the first root of $j_{n}(z)$, with $n=1$,
 to describe low aspect ratio high $\beta$
 toroidal plasma equilibria.

\newpage
\section{MAGNETIC AND CURRENT STRUCTURES}

With the spatial structure solved,
 the magnetic field components are given by

\begin{mathletters}
\begin{eqnarray}
\label{eqno23a}
B_{r}\,=\,+{1\over r\sin\theta}
 {1\over r}{\partial P\over\partial\theta}\,
 =\,-{1\over r^2}R(r)
 {d\Theta(x)\over dx}\,\,\,,
\\
\label{eqno23b}
B_{\theta}\,=\,-{1\over r\sin\theta}
 {\partial P\over\partial r}\,
 =\,-{1\over r}{dR(r)\over dr}
 {1\over (1-x^2)^{1/2}}\Theta(x)\,\,\,,
\\
\label{eqno23c}
B_{\phi}\,=\,+{1\over r\sin\theta} Q(P)\,\,\,.
\end{eqnarray}
\end{mathletters}

\noindent The solution $R(r)$ vanishes at some $r$
 where we have $B_{r}(r)=0$.
 The solution $\Theta(x)$ also vanishes at some $x$.
 Together they describe the magnetic fields.
 Within this region of $(r,x)$, the topological center
 defined by $dR(r)/dr=0$ and $d\Theta(x)/dx=0$
 has $B_{r}=0$ and $B_{\theta}=0$.
 This is the magnetic axis, $r=r_{*}$,
 where the magnetic field is entirely toroidal.
 The field lines about this center are given by

\begin{equation}
\label{eqno24}
{B_{r}\over dr}\,=\,{B_{\theta}\over rd\theta}\,
 =\,{B_{\phi}\over r\sin\theta d\phi}\,\,\,.
\end{equation}

\noindent By axisymmetry, the third group
 is decoupled from the first two groups.
 For the field lines on an $(r-\theta)$ plane,
 we consider the first equality between $B_{r}$ and $B_{\theta}$
 which gives

\begin{equation}
\label{eqno25}
P(r,x)\,=\,R(r)\Theta(x)\,=\,C\,\,\,.
\end{equation}

\noindent The nested poloidal field lines
 are given by the contours of $P(r,x)$ on the $(r-x)$ plane.
 At the topological center, we have $\Theta(x)$ maximum and
 $R(r)$ maximum, so that $P(r,x)$ is maximum.
 Since $r\sin\theta$ is the distance
 of a point on the $(r-x)$ plane to the z axis,
 Eq.~\ref{eqno23c} states that the line integral
 of $B_{\phi}$ around the circle on the azimuthal plane
 is measured by $2\pi Q$,

\begin{eqnarray}
\nonumber
2\pi r\sin\theta B_{\phi}\,=\,2\pi Q\,=\,\mu I_{z}\,\,\,.
\end{eqnarray}

\noindent This line integral about the axis of symmetry
 is maximum at the topological center.
 Also, it is evident that $Q$ is equivalent to the axial current,
 where the constant part $Q_{0}$ amounts to a uniform component.
 As for $P$, we evaluate the poloidal magnetic flux
 by integrating Eq.~\ref{eqno23b} on the $x=0$ plane
 over a cross section to give
 
\begin{eqnarray}
\nonumber
\int_{r_{*}}^{r} 2\pi r B_{\theta}dr\,
 =\,-2\pi (P(z)-P(z_{*}))\,\,\,.
\end{eqnarray}

\noindent As for the current density of Eq.~\ref{eqno8},
 making use of the Grad-Shafranov equation
 of Eq.~\ref{eqno15} gives
  
\begin{eqnarray}
\label{eqno26}
\mu\vec J\,
 =\,{A_{0}\over r\sin\theta}
 \{+{1\over r}{\partial Q\over\partial\theta},
   -{\partial Q\over\partial r},
   +(a^2P-{1\over\alpha^{2}}({v_{s}\over A'_{0}})^{2}r^2\sin^2\theta
 {\partial\over\partial P}\ln (1-\mu\rho_{0}g^{2}\rho(P))\}
 \,\,\,.
\end{eqnarray}

\noindent Analogous to the magnetic field lines,
 the current density field lines are given by
 
\begin{equation}
\label{eqno27}
{J_{r}\over dr}\,=\,{J_{\theta}\over rd\theta}\,
 =\,{J_{\phi}\over r\sin\theta d\phi}\,\,\,.
\end{equation}

\noindent Considering the first equality,
 the poloidal current density contours are given by

\begin{eqnarray}
\label{eqno28}
Q(r,x)\,=\,C\,\,\,.
\end{eqnarray}

\newpage
\section{L AND H MODES}

We note that there are two independent solutions
 for $R_{0}(r)$ in Eq.~\ref{eqno21}.
 The first one is $zj_{1}(z)$
 which vanishes at $z=0$.
 With $z_{1}$ and $z_{2}$ as the first and second zeros,
 the region bounded by $0<z<z_{1}$
 could be used to describe spheromak and
 high $\beta$ low aspect ratio tokamak equilibria.
 The second one is $zy_{1}(z)$
 which diverges at $z=0$.
 Since our domain of interest in tokamak plasmas
 excludes $z=0$,
 the singularity of $y_{1}$ is irrelevant.
 The region bounded by $z_{1}<z<z_{2}$
 could also be used to describe
 high $\beta$ low aspect ratio tokamak equilibria.
 The functions $zj_{1}(z)$ and $zy_{1}(z)$ are shown in Fig.1.
 The poloidal magnetic contours of Eq.~\ref{eqno25}
 for $zj_{1}(z)$ in the interval
 $0<z<z_{1}$ are shown in Fig.2.
 In particular, this solution
 could also be applied to spheromaks
 where $z=0$ is accessible to plasma equilibria.
 Similar contours for $z_{1}<z<z_{2}$,
 $z_{2}<z<z_{3}$, and so on, can be obtained
 to represent tokamak plasmas of different aspect ratios. 
 The contours for $zy_{1}(z)$ in the interval
 $z_{1}<z<z_{2}$ are shown in Fig.3.
 In order to illustrate the essential features,
 we have neglected the particular solution $R_{1}(r)$,
 and have taken $R(r)=R_{0}(r)$.
 The contour levels are taken at
 $0.95, 0.9, 0.7, 0.5, 0.3, 0.1$ of the respective peak value.
 The external contours indicate high poloidal fields,
 and internal contours for low poloidal fields.
 These contours also indicate
 the poloidal rotations with rotation velocity
 high on the outside and low on the inside.

Bounded by a smaller interval $z_{1}<z<z_{2}$,
 we note that Fig.3 of $zy_{1}(z)$
 has a more localized domain and steeper edge profile
 than the equilibrium in Fig.2,
 which is described by $zj_{1}(z)$
 in the larger interval $0<z<z_{1}$.
 Including the negative valued particular solution
 $R_{1}(r)$ further steepens the edge gradient.
 We associate $zj_{1}(z)$ and $zy_{1}(z)$
 to the L and H mode respectively.
 We also note that the poloidal and toroidal magnetic fields
 are plotted in normalized radial coordinate $z=ar$.
 In laboratory plasmas,
 the fields are measured in terms of radius $r$.
 To connect to our normalized results,
 we need to determine the normalizing parameter $a$.
 This can be done by considering
 the magnetic axis $r_{*}$ of a laboratory plasma,
 say in the $zj_{1}(z)$ mode bounded by $0<z<z_{1}$,
 through $a_{j}r_{*}=z_{*}=2.7$.
 Defined by the divertor scrape-off,
 the radial range of laboratory plasma, $r_{a}<r<r_{b}$, 
 can now be converted to $0<z_{a}<z<z_{b}<z_{1}$
 with $z_{1}=4.5$.
 In the case of $zy_{1}(z)$ mode, we have $a_{y}r_{*}=z_{*}=4.5$,
 giving $a_{y}/a_{j}=4.5/2.7$.
 The radial range can be converted to
 $z_{1}<z_{a}<z<z_{b}<z_{2}$ with $z_{1}=2.8$ and $z_{2}=6.1$.
 Experimentally, this change of the normalizing parameter
 from $a_{j}$ to $a_{y}$ could be accomplished
 by pellet injection and high power external heating.

In the L mode, the $zj_{1}(z)$ profile is more diffused
 between a larger interval of the zeros.
 The divertor action removes the edge plasma
 to the $z_{a}<z<z_{b}$ domain,
 with poloidal rotation velocity contours
 corresponding to such domain.
 In the H mode, due to the compactness of the interval
 between zeros of $zy_{1}(z)$ profile,
 the plasma equilibrium fits within the toroidal machine vessel
 naturally with much less divertor shaping.
 The plasma equilibrium occupies
 probably the entire domain $z_{1}<z<z_{2}$,
 or a large part of it.
 As a result, poloidal rotation contours
 of the H mode cover not just the central part
 but also the high velocity part on the outside.
 By going from L to H mode,
 the rotation contours within the plasma cross-section,
 defined by the divertor action,
 are enlarged from a partial central profile
 to an almost complete profile.
 Observed at a fixed position at the plasma edge,
 we would have the impression
 that the rotation velocity has been speeded up.

The corresponding mass density contours of Eq.(17b)
 for L and H modes are shown in Fig.4 and Fig.5 respectively.
 The profiles along $x=0$ horizontal cut are shown in Fig.6.
 Although the two modes are presented in one same figure
 showing approximately the same dimensionless amplitudes,
 the physical amplitude and dimension
 is given by $\rho_{0}$ defined in Eq.~\ref{eqno10}.
 As a result, the mass density of the H mode
 could be much larger than that of the L mode.
 The essence of Fig.6 is to show the relative shape
 of the mass density profiles for the two modes.
 We have suggested the identification
 of L mode to the $zj_{1}(z)$ solution in the $(0,z_{1})$ domain,
 and H mode to the $zy_{1}(z)$ solution in the $(z_{1},z_{2})$ domain.
 To map these $z$ domains to the same $r$ domain of machine vessel,
 We have used two different normalizing parameters
 $a_{j}$ and $a_{y}$ for the source function $Q^{2}(P)$
 of Eq.~\ref{eqno17a}.
 Since $a_{y}=1.7a_{j}$, this would require
 a substantial increase of toroidal magnetic field
 according to Eq.~\ref{eqno23c}.
 To avoid this substantial toroidal field enhancement,
 the H mode could be generated by superimposing
 the $zy_{1}(z)$ solution to the $zj_{1}(z)$ solution,
 without displacing significantly the $z$ domain.
 As an example, with $\lambda_{0}=1$,
 the profile of $(zj_{1}(z)+zy_{1}(z))$ is shown
 in Fig.7 indicating $z_{1}=1.9$ and $z_{2}=5.3$
 with a maximum at $z_{*}=3.6$.

\newpage
\section{DISCUSSIONS AND CONCLUSIONS}

Experimentally, due to the divertor action on the edge plasma,
 there is an electrostatic field normal to the magnetic surfaces.
 This field is particularly large in the H mode configuration
 because of the steep edge gradient.
 Since the poloidal field lines are described by the $P$ contours,
 this field can be written as $\vec E=-\nabla\Phi(P)$,
 which warrants $\partial\vec B/\partial t=0$ for equilibrium.
 Interacting with magnetic islands on a rational surface,
 this electric field drives zonal flows
 that establish transport barriers
 for better plasma density and energy confinements.
 With the rotational toroidal equilibrium approach,
 the L/H transition amounts to a bifurcation
 of the L equilibrium to the H equilibrium,
 under the actions of external pumping
 through pellet injection and strong auxiliary heating
 for density and temperature profile shaping.
 The normal electric field, zonal flows, and transport barriers,
 come as consequences, not cuases,
 of the steep edge gradient of the H mode.
 
We have solved toroidal plasma equilibria
 with axisymmetric toroidal and poloidal rotations,
 that are self-similar to the corresponding magnetic fields.
 The rotational Grad-Shafranov equation
 in spherical coordinates is solved for toroidal solutions,
 under the assumption that the scalar rotational pressure
 is much less than the plasma pressure.
 With a specific set of source functions,
 there are two independent homogeneous radial modes
 given by $zj_{1}(z)$ and $zy_{1}(z)$.
 The $zj_{1}(z)$ mode in the region $0<z<z_{1}$ and
 the $zy_{1}(z)$ mode in the region $z_{1}<z<z_{2}$
 could be applied to current large scale tokamaks.
 The $zj_{1}(z)$ mode has a diffuse edge profile
 over a larger $z$ domain,
 and the $zy_{1}(z)$ mode has a steep edge profile
 over a smaller $z$ domain.
 We associate them to the L and H modes respectively.
 The L/H transition amounts to a bifurcation
 of one equilibrium configuration to another,
 following a change of the normalizing parameter
 from $a_{j}$ to $a_{y}$.
 Experimentally, this change of parameter could be achived
 by pellet injection and large external heating.
 

{}

\clearpage
\begin{figure}
\plotone{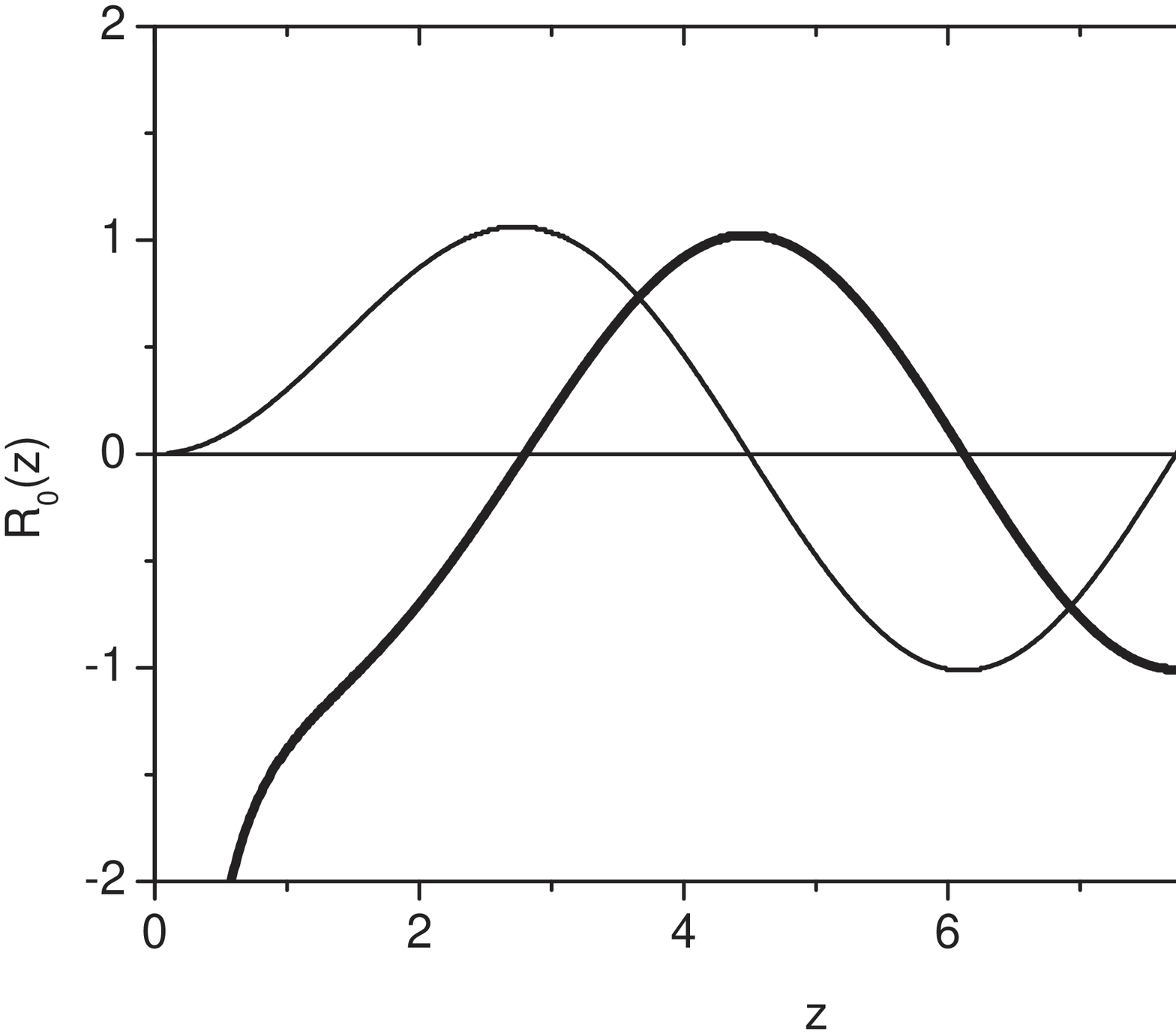}
\caption{The functions $zj_{1}(z)$, in thin line,
 and $zy_{1}(z)$, in thick line,
 of the two independent homogeneous
 $R_{0}(z)$ radial solutions
 are plotted as a function of $z$.}
\end{figure}

\clearpage
\begin{figure}
\plotone{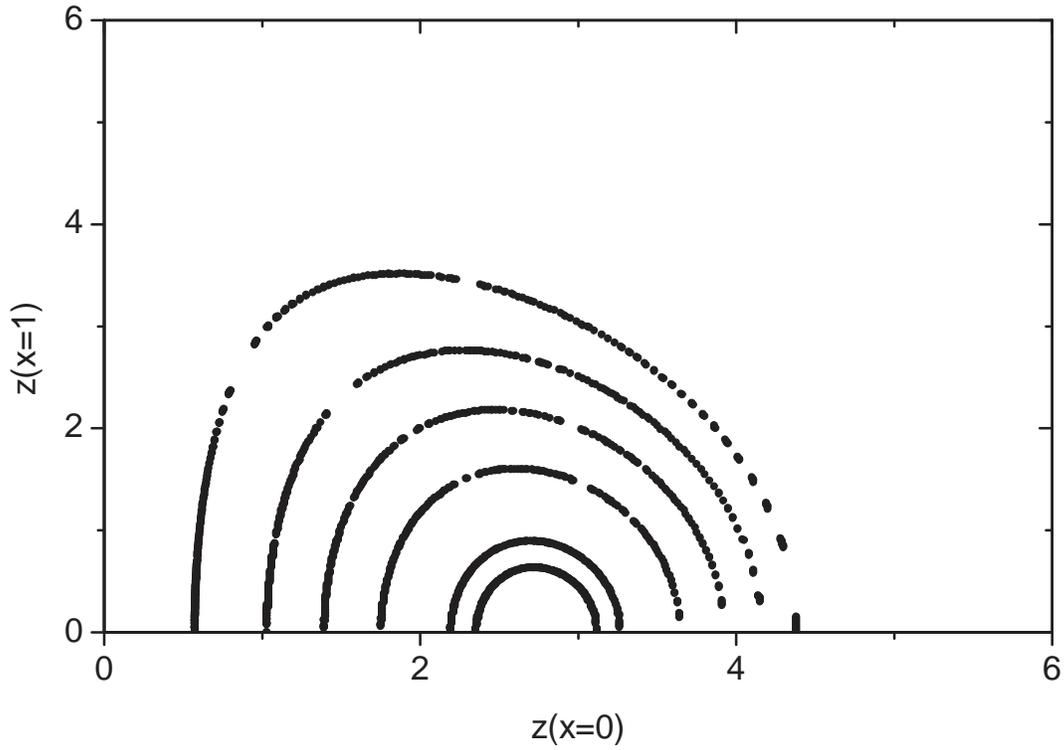}
\caption{The poloidal magnetic field lines
 of $R_{0}(z)=zj_{1}(z)$ L mode
 are shown in polar plots
 with increasing contour labels
 from outer to inner contours.
 These contours are shared by the poloidal rotation.
 The axes are labelled in $z$ together with $x=\cos\theta$.}
\end{figure}

\clearpage
\begin{figure}
\plotone{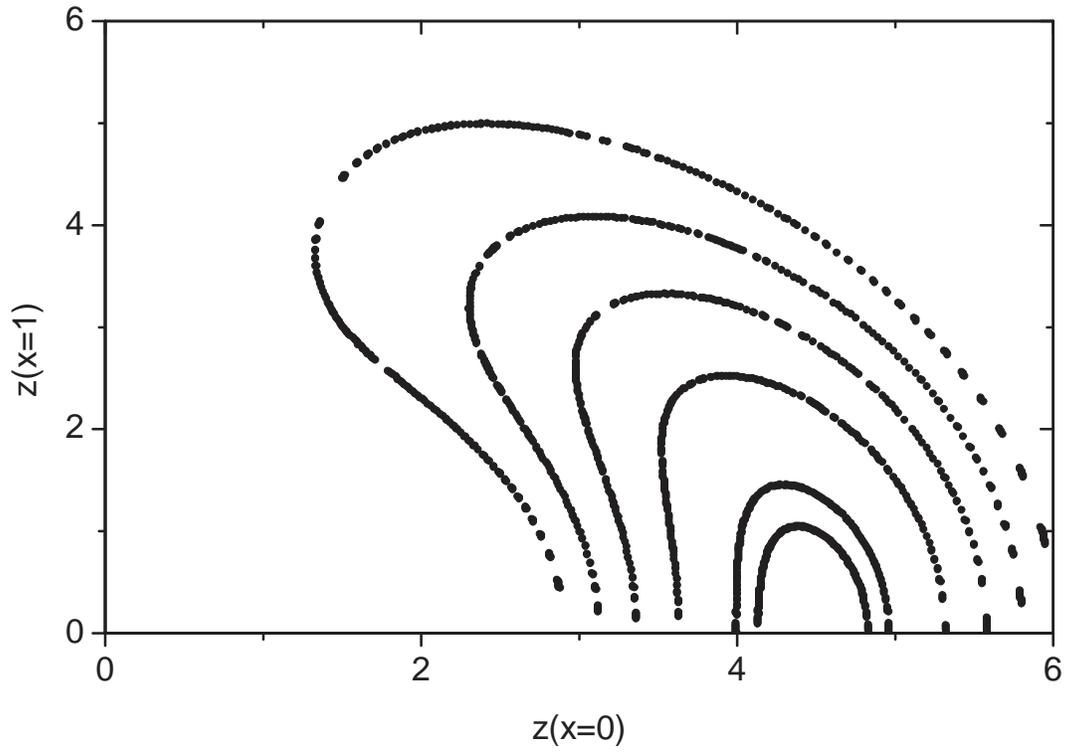}
\caption{The poloidal magnetic field lines
 of $R_{0}(z)=zy_{1}(z)$ H mode
 are shown in polar plots
 with increasing contour labels
 from outer to inner contours.}
\end{figure}

\clearpage
\begin{figure}
\plotone{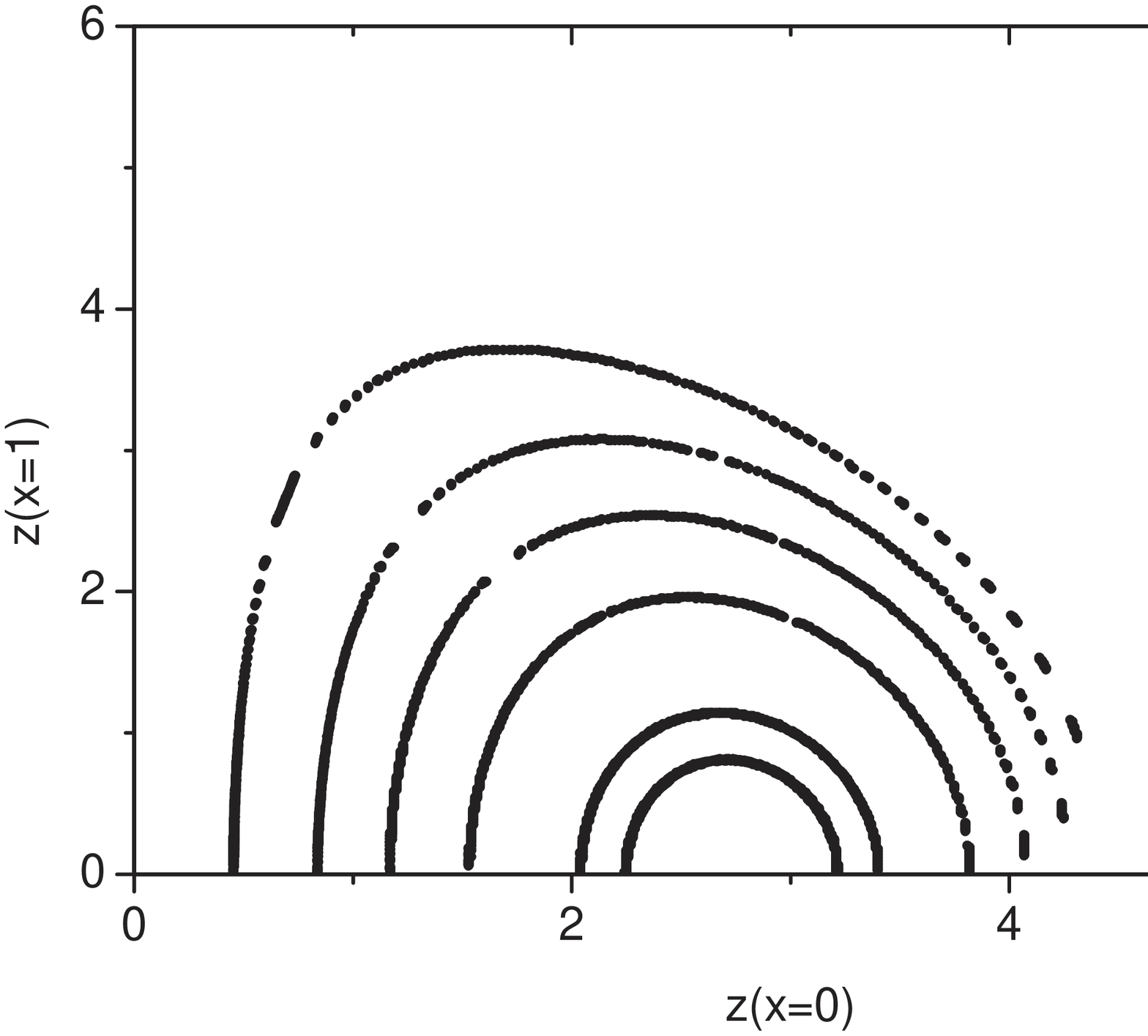}
\caption{The mass density contours
 of $R_{0}(z)=zj_{1}(z)$ L mode
 are shown in polar plots
 with increasing contour labels
 from outer to inner contours.}
\end{figure}

\clearpage
\begin{figure}
\plotone{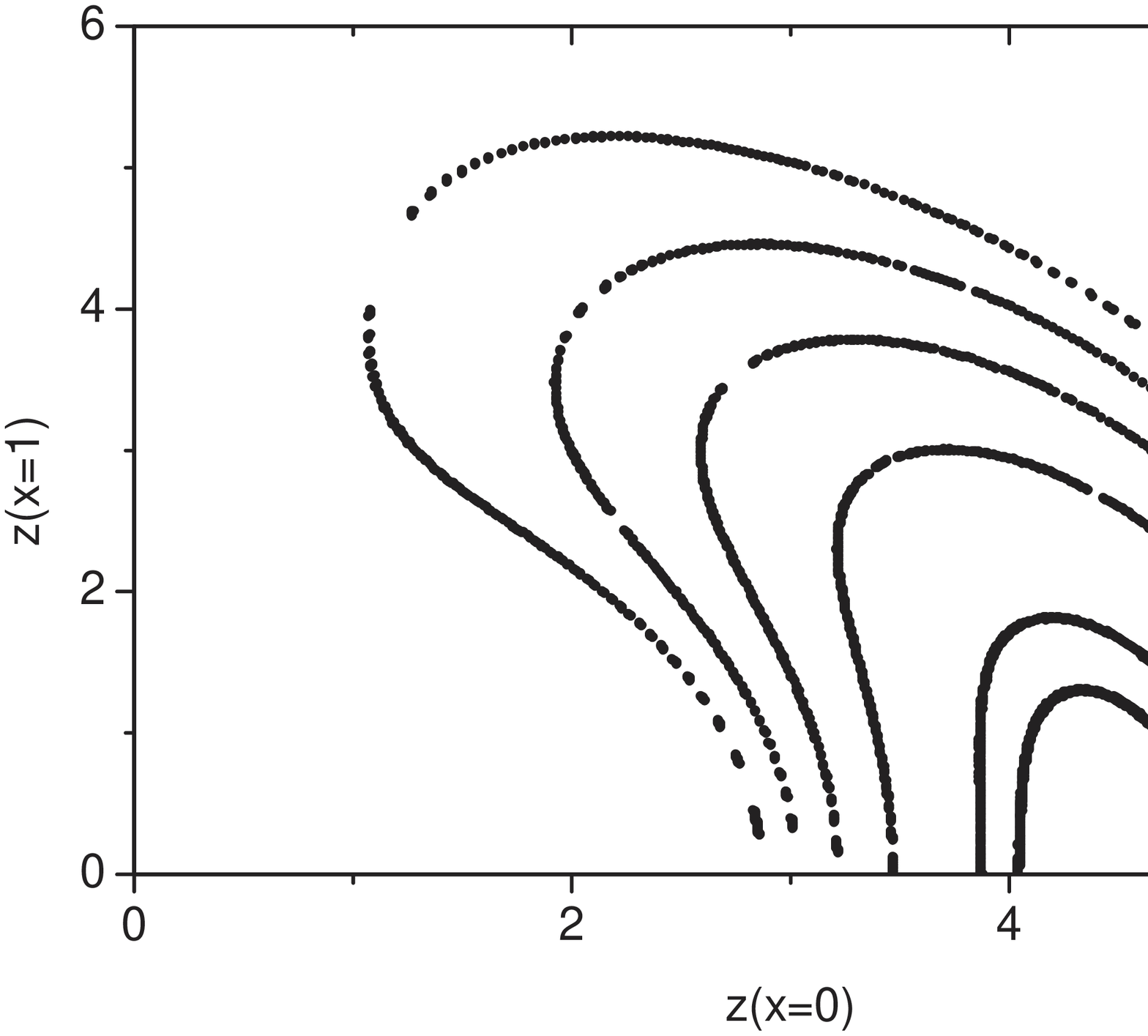}
\caption{The mass density contours
 of $R_{0}(z)=zy_{1}(z)$ R mode
 are shown in polar plots
 with increasing contour labels
 from outer to inner contours.}
\end{figure}
\clearpage

\begin{figure}
\plotone{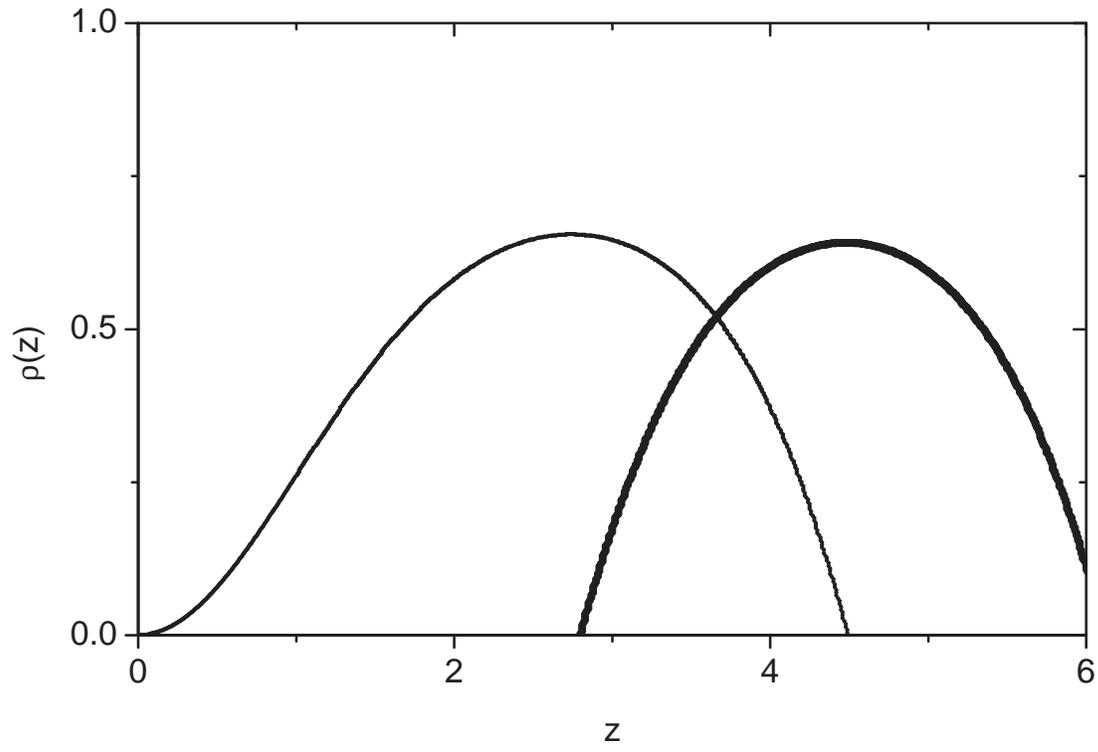}
\caption{The mass densities along $x=0$
 for $zj_{1}(z)$ mode, in thin line,
 and $zy_{1}(z)$ mode, in thick line,
 are plotted as a function of $z$.}
\end{figure}

\begin{figure}
\plotone{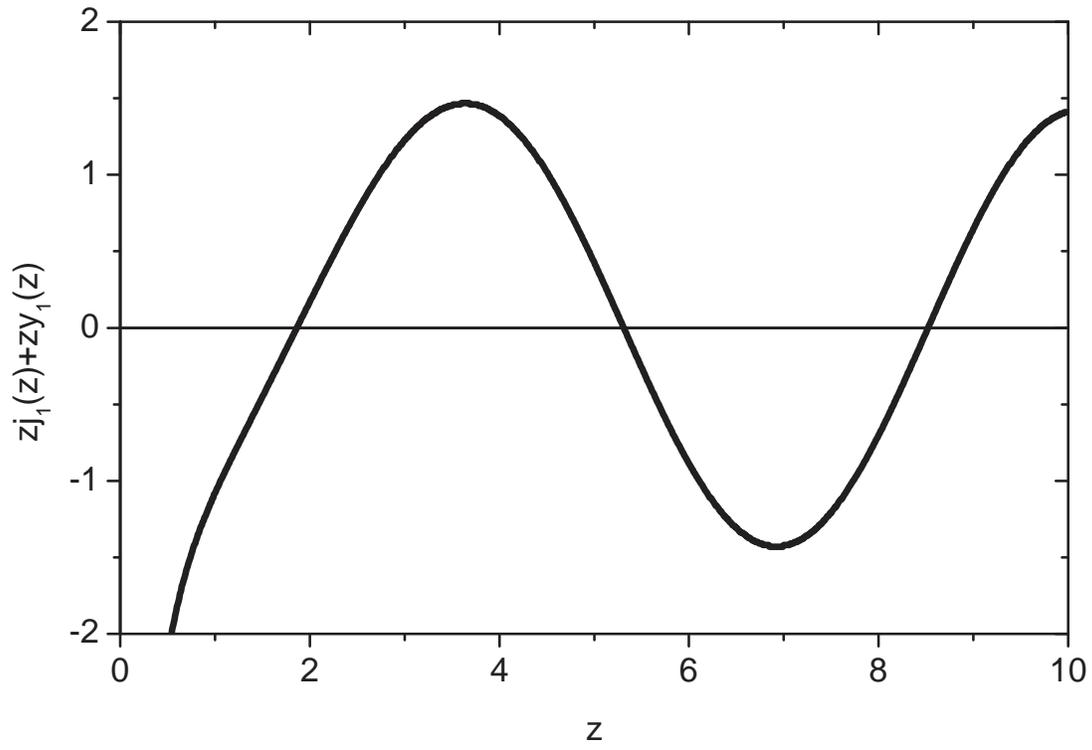}
\caption{The function $zj_{1}(z)+zy_{1}(z)$
 is plotted as a function of $z$
 indicating $z_{1}=1.9$ and $z_{2}=5.3$.}
\end{figure}

\end{document}